\documentclass[a4paper,twocolumn,english,aps,prl,reprint,superscriptaddress,showpacs,showkeys,longbibliography]{revtex4-1}
\usepackage[latin9]{inputenc}
\usepackage{babel}
\usepackage{mathtools}
\usepackage{amsmath}
\usepackage{amssymb}
\usepackage{graphicx}
\usepackage{float}
\usepackage[colorlinks=true, linkcolor=red, citecolor=blue, urlcolor=blue]{hyperref}
\usepackage[usenames,dvipsnames]{color}
\usepackage{bm}
\usepackage{bbold}
\usepackage{braket}
\usepackage{comment}
\usepackage{dcolumn,extarrows}
\usepackage{soul}
\usepackage{footmisc}
\usepackage{pdfpages}
\usepackage{pgf,tikz}

\definecolor{mygreen}{rgb}{0,0.5,0}
\definecolor{myblue}{rgb}{0,0,0.75}
\definecolor{mymagenta}{cmyk}{0,1,0,0.12}

\definecolor{mygreen}{rgb}{0,0.5,0}\definecolor{myblue}{rgb}{0,0,0.75}\definecolor{mymagenta}{cmyk}{0,1,0,0.12}

\newcommand{\bla}[1]{\left( #1 \right)}
\newcommand{\blb}[1]{\left[ #1 \right]}

\makeatletter
\AtBeginDocument{\let\LS@rot\@undefined}
\makeatother

\begin{document}

\title{Observation of Floquet Raman transition in a driven solid-state spin system}
\author{Zijun Shu}
\thanks{These authors contributed equally to this work.}
\author{Yu Liu}
\thanks{These authors contributed equally to this work.}
\author{Qingyun Cao}
\author{Pengcheng Yang}
\author{Shaoliang~Zhang}
\email{shaoliang@hust.edu.cn}
\affiliation{School of Physics, Huazhong University of Science and Technology, Wuhan 430074, China}
\affiliation{International Joint Laboratory on Quantum Sensing and Quantum Metrology,
Huazhong University of Science and Technology, Wuhan, 430074, China}
\author{Martin B. Plenio}
\affiliation{Institut f\"{u}r Theoretische Physik $\&$ IQST, Albert-Einstein Allee 11, Universit\"{a}t Ulm, D-89081 Ulm, Germany}
\affiliation{International Joint Laboratory on Quantum Sensing and Quantum Metrology,
Huazhong University of Science and Technology, Wuhan, 430074, China}
\author{Fedor Jelezko}
\affiliation{Institut f\"{u}r Quantenoptik $\&$ IQST, Albert-Einstein Allee 11, Universit\"{a}t Ulm, D-89081 Ulm, Germany}
\affiliation{International Joint Laboratory on Quantum Sensing and Quantum Metrology,
Huazhong University of Science and Technology, Wuhan, 430074, China}
\author{Jianming Cai}
\email{jianmingcai@hust.edu.cn}
\affiliation{School of Physics, Huazhong University of Science and Technology, Wuhan 430074, China}
\affiliation{International Joint Laboratory on Quantum Sensing and Quantum Metrology,
Huazhong University of Science and Technology, Wuhan, 430074, China}

\begin{abstract}
We experimentally observe Floquet Raman transitions in the weakly driven solid state
spin system of nitrogen-vacancy center in diamond. The periodically driven spin system simulates
a two-band Wannier-Stark ladder model, and allows us to observe coherent spin state transfer
arising from Raman transition mediated by Floquet synthetic levels. It also leads to
the prediction of analog photon-assisted Floquet Raman transition and dynamical localisation in a 
driven two-level quantum system. The demonstrated rich Floquet
dynamics offers new capabilities to achieve effective Floquet coherent control of a
quantum system with potential applications in various types of quantum technologies based
on driven quantum dynamics. In particular, the Floquet-Raman system may be used as a quantum simulator for the physics of periodically driven systems.
\end{abstract}


\date{\today}

\maketitle

Coherent control of quantum system is an essential prerequisite in a wide 
range of quantum experiments \cite{Gordon97,Shapiro2000}. In particular, together with quantum initialization and readout, it represents an important ingredient for the realization of quantum technology, including quantum computing \cite{Nielsen10}, quantum simulation \cite{Feynman82,Lloyd96} and quantum sensing \cite{Degen_RMP2017}. Time-dependent 
periodic driving, as implemented for example by laser and microwave field, is commonly used for coherent 
quantum control of atom and spin systems \cite{Shirley1965, Chuang_RMP2005,Blatt_Nature2008}. For two-level quantum systems, the well known Rabi oscillation induced by a weak periodic driving field on resonance leads to coherent oscillatory state transitions. Stimulated Raman transitions between 
two ground states via a far detuned third excited state provide an extremely powerful tool 
for coherent manipulation \cite{Shore90}. Moreover, periodic driving is an useful tool that may lead to
new exotic phases, such as topological Floquet insulators \cite{Kitagawa2010,Rudner13,Rechtsman13,Lindner11,Rudner13,Abanin16,Wang15} and 
time crystals \cite{Yao2014,Choi17,Zhang17,Martin17}. An important consequence of periodic driving is 
the emergence of synthetic dimensions supported by Floquet dressed state which opens the field
of Floquet Hamiltonian engineering \cite{Goldman_PRX14} and new techniques for coherent quantum control
\cite{Deng15}.

Periodic driving is not only the basis of coherent quantum control \cite{Shirley1965}, it also plays an important role in several fundamental concepts in quantum dynamics, e.g. geometric phase \cite{Berry_84} and adiabatic quantum algorithm, where the validity of adiabatic evolution \cite{Ehrenfest1916,Kato1936,Born1928,Zurek05,Brooke99,Farhi01} is usually required. The traditional adiabatic condition requires the change of Hamiltonian to be sufficiently slow on the time scale that is proportional to the inverse squared energy gap. Such an adiabatic condition was however found to be neither sufficient nor necessary to guarantee adiabaticity \cite{Marzlin04,Tong10,Tong05,Li2014}. An observation of non-adiabaticity violating the traditional adiabatic condition was shown in \cite{Du2008PRL} which can be explained by Rabi resonance. The general anomalous non-adiabaticities appearing in a weakly periodically driven two-level system has fundamental connection with Floquet resonance \cite{Russomanno17}. The experimental demonstration of general anomalous non-adiabaticities in periodically driven systems remains challenging due to the requirement of sufficiently long coherence time and precise quantum control.

In this work, we experimentally study Floquet dynamics of a single nitrogen-vacancy (NV) center spin in diamond driven by a {\it weak} microwave field. As compared with strong driving \cite{Deng15}, the driving strength is much weaker than the frequency detuning and conventionally it would not induce spin state transition. With such a weakly driven two-level system, we simulate a two band Wannier-Stark ladder model \cite{Grosso2000,Mendez81}, and instead observe coherent oscillation between spin states. The phenomenon represents Raman transition between Floquet synthetic levels \cite{Russomanno17} albeit the system itself is a two-level quantum system. We also show that analog photon-assisted Raman transition can be realised among Floquet levels, which provides a new knob for the engineering of quantum control. We expect that the present idea can be extended to simulate and investigate non-trivial effects of a Wannier-Stark ladder model, e.g. with complex hopping amplitudes, by exploiting Floquet synthetic levels. In addition, the observed Floquet Raman transitions manifest general anomalous non-adiabaticities in periodically driven system, and may help to gain further insight into adiabatic quantum dynamics.
%

%
\begin{figure}[t]
\begin{center}
\hspace{-0.0cm}
\includegraphics[width=0.98\linewidth]{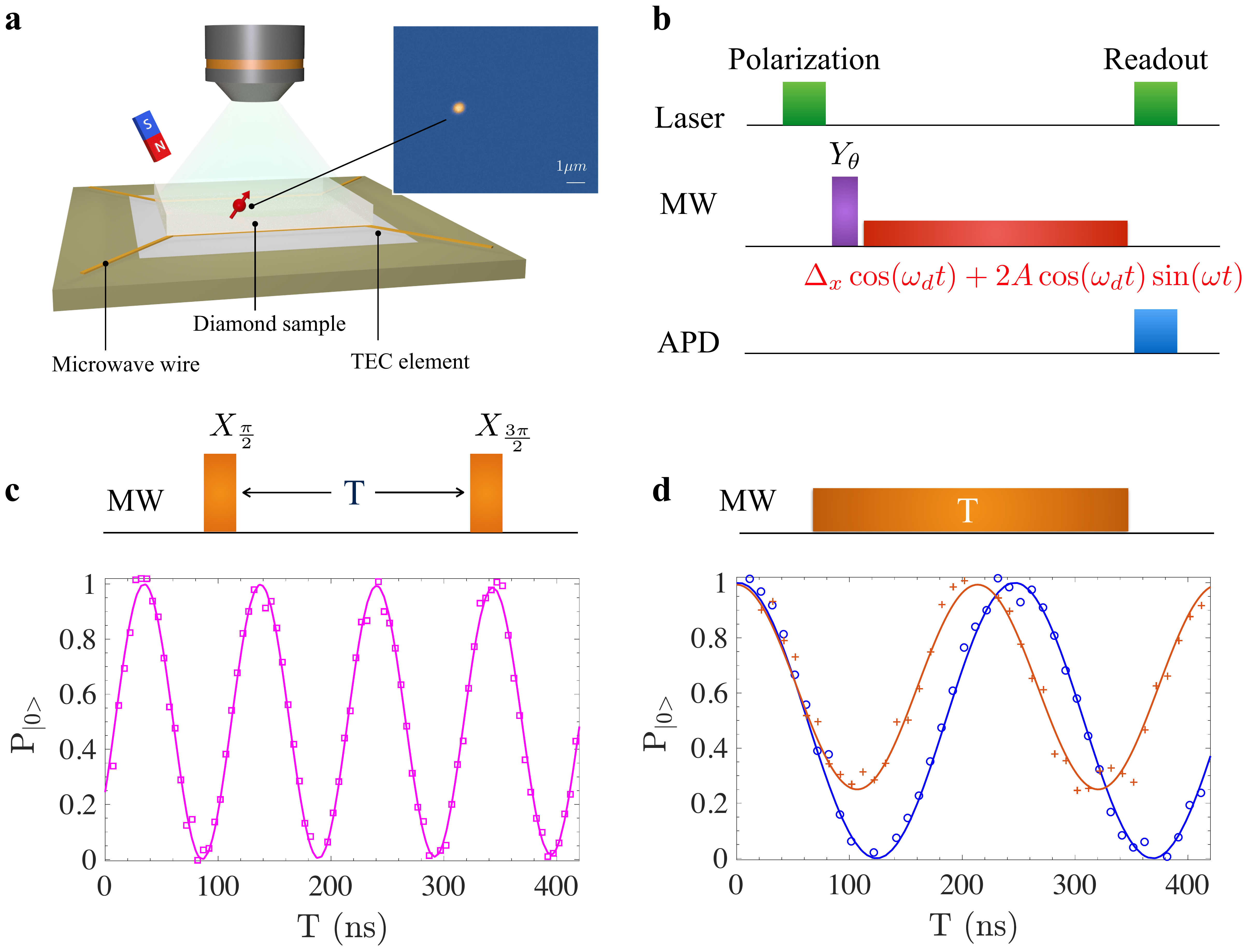}
\end{center}
\caption{(Color online) {Setup for observation of Floquet Raman transition of NV center spin in diamond}. ({\bf a}) Schematic representation of the experimental setup. The temperature of the diamond sample is stabilized by a TEC element. The NV center spin is coupled to the microwave driving field via a copper wire waveguide. The readout of the NV center spin state is achieved by spin-dependent fluorescence. Confocal scan image of an NV center is shown in the inset. ({\bf b}) Pulse sequence for Floquet Raman transition. MW corresponds to the microwave field, and the optical measurement is performed by a photodetector (APD). The initial spin state is prepared with a microwave pulse $Y_{\theta}$ (rotation around $\hat{y}$ axis by angle $\theta$), the duration of which is $ \tau=\theta/2\Omega$ where $\Omega$ is the amplitude of the microwave pulse. The synthesized microwave driving field $V(t)=\Delta_x \cos(\omega_d t) + 2A \cos(\omega_d t) \sin\bla{\omega t} $ implements Floquet quantum control. ({\bf c}) shows Ramsey measurement, the oscillation of which allows us to determine the frequency detuning $\Delta_z$. ({\bf d}) shows Rabi oscillations with two driving frequencies $(\omega_d/2\pi)=1445.8$ MHz (blue, $\circ$) and $1443.8$ MHz (red, $+$). Measurement of the effective Rabi frequency $\omega_0=\bla{\Delta_z^2+\Delta_x^2}^{1/2}$ gives an estimation of the driving amplitude $\Delta_x=(2\pi)4.06\mbox{MHz}$. }
\label{fig:setup}
\end{figure}
%

The two-level quantum system in our experiment is a spin-$\frac{1}{2}$ system from the subspace of the triplet ground state manifold of a single NV center in electronic grade bulk diamond (with less than 5 ppb nitrogen impurities). The scan image of NV center as shown in Fig.\ref{fig:setup}(a) is obtained from a home-built confocal setup. The NV center spin has the advantage of long coherence time (i.e. narrow line broadening) under ambient condition, which is mainly dependent on the nuclear spin bath in diamond and temperature fluctuation for the present sample. The NV center has a triplet ground state with three spin sublevels $m_s=0,\pm 1$. The degeneracy of $m_s=\pm 1$ is lifted by applying an external magnetic field $B_z$ along the NV axis, which provides an effective two-level system supported by the ground state sublevels $m_s=0$ and $m_s=-1$ with an energy gap $\Delta E= D -\gamma B_z$, where the zero field splitting is $D=(2\pi) 2.87\mbox{GHz}$ and $\gamma$ is the electron gyromagnetic ratio. We apply a magnetic field $B_z=509$ G along the NV axis so that the nitrogen nuclear spin is polarized. The value of $\Delta E$ is determined by pulsed optically detected magnetic resonance measurement (pulsed ODMR), which also confirms the polarization of the nitrogen nuclear spin \cite{SI}. The stability of the energy splitting is very critical to observe Floquet Raman transition \cite{Russomanno17}. In our experiment setup, we mount the diamond sample on a single-stage TEC element under temperature control. The temperature of the diamond sample is stabilized by a 12W temperature controller (Thorlabs TED200C) so that the temperature fluctuation is reduced down to $0.1$K that corresponds to a line broadening of $\sim 7.7$ kHz \cite{Acosta10}. The Ramsey measurement suggests a line broadening of $\sim 40 \mbox{kHz}$ for the NVs used in our experiment, which corresponds to a coherence time of $T_2^*  \simeq 4\mu s$, see Fig.\ref{fig:setup}(c) where the $m_s$=0 state probability $P_{\ket{0}}$ is calibrated from the florescence data as normalized by resonant Rabi oscillation.

%
\begin{figure*}
\begin{center}
\hspace{-0.2cm}
\includegraphics[width=0.74\linewidth]{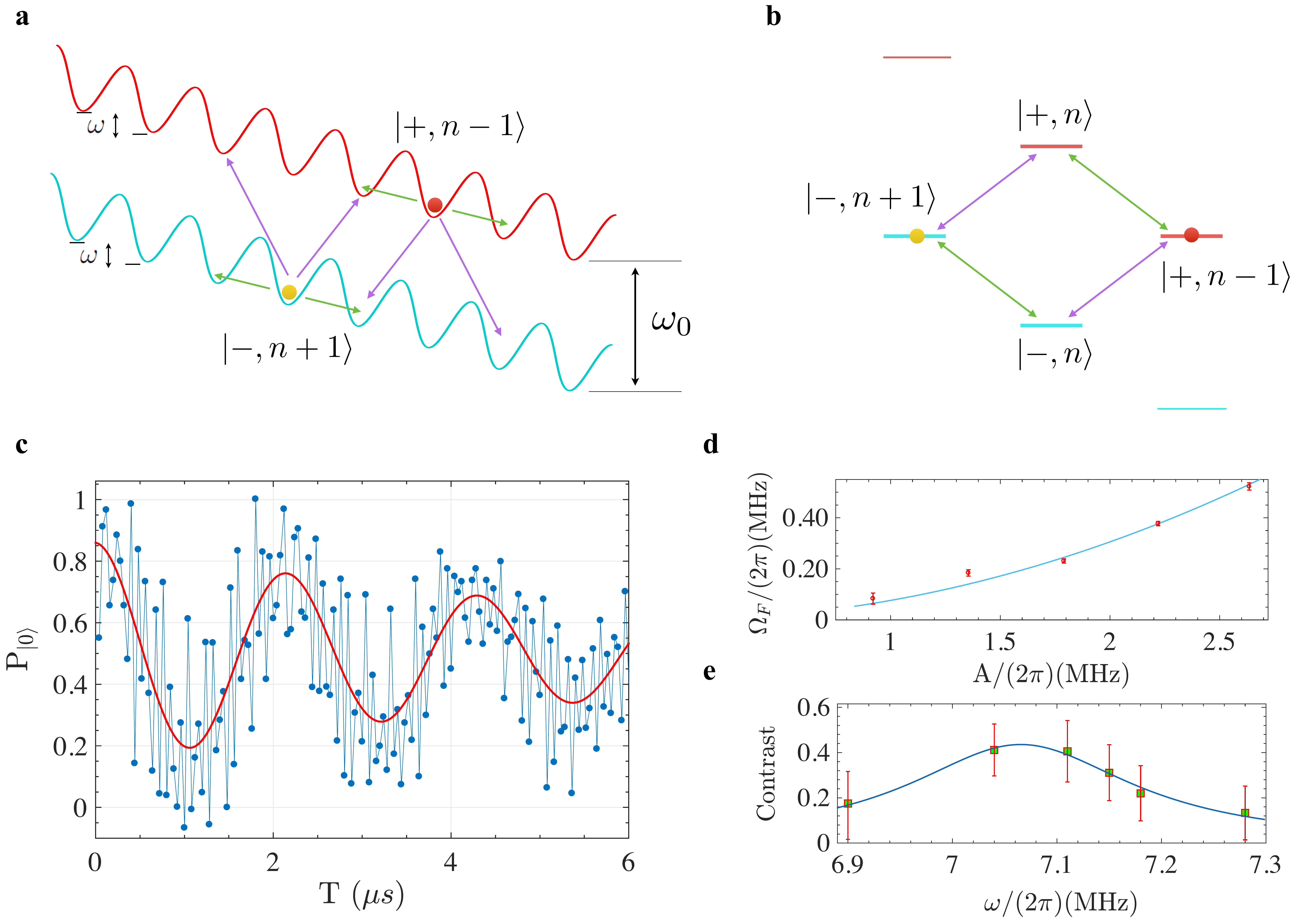}
\end{center}
\caption{(Color online) Floquet Raman transition of NV center spin in diamond. ({\bf a}) A two band Wannier-Stark ladder model is simulated by a periodically driven spin-$\frac{1}{2}$ system. The energy gap between the upper and lower band is $\omega_0$, and the ladder energy spacing is given by the frequency $\omega$ of the weak driving field. The arrows represent couplings between Floquet synthetic levels. ({\bf b}) Raman transition between the lower and upper band levels $\ket{-,n+1}$ and $\ket{+,n-1}$ is mediated by Floquet synthetic levels $\ket{+,n}$ and $\ket{-,n}$. ({\bf c}) Coherent spin state transfer is realized by Floquet Raman transition. The red curve represents inter-band transition filtering out the fast oscillating component. The strength and frequency of the weak microwave driving field are $A=(2\pi)  2.37$ MHz and $\omega=(2\pi)  6.985$ MHz. The other parameters are $\Delta_z=(2\pi) 10.03$ MHz, and $\Delta_x=(2\pi) 9.67 $ MHz. ({\bf d}) The dependence of Floquet Raman transition Rabi frequency $\Omega_F$ on the driving strength $A$, as compared with the numerical simulation result (cyan curve). The parameters are $\omega=(2\pi) 7.09 $ MHz, $\Delta_z=(2\pi) 9.92$ MHz, and $\Delta_x=(2\pi) 10.12$ MHz. ({\bf e}) The contrast of Floquet Raman transition as a function of the driving frequency $\omega$. The parameters are $A=(2\pi) 1.37$ MHz, $\Delta_z=(2\pi) 9.63$ MHz, and $\Delta_x=(2\pi) 10.32$ MHz. The solid curve is a fit by a Lorentzian function $\sim \gamma^2/\blb{\gamma^2+(\omega-\omega_0 )^2 }$.
}
\label{fig:frt}
\end{figure*}
%

%
To implement Floquet quantum control and observe Floquet Raman transition, we synthesize a microwave field described by $V(t)=\Delta_x \cos(\omega_d t) + 2A \cos(\omega_d t) \sin\bla{\omega t} $ with  $\Delta_x, A \ll \omega_d $ with accurate timing and amplitude by using a Tektronics arbitrary waveform generator. In the interaction picture, we find the effective Hamiltonian for the two-level system \cite{SI}
\begin{equation}
H=\frac{\Delta_z}{2}\sigma_z+\frac{\Delta_x}{2}\sigma_x+A\sin\bla{\omega t}\sigma_x,\label{eq:Hamiltonian}
\end{equation}
where we set $\hbar=1$, $\Delta_z=\Delta E-\omega_d$, $\sigma_x, \sigma_z$ are Pauli operators for the spin-$\frac{1}{2}$ system, $\Delta_x$ and $\Delta_z$ represent the transversal and longitudinal components of the energy splitting. In our experiment, we exploit Floquet synthetic levels supported by the periodically driven Hamiltonian in Eq.(\ref{eq:Hamiltonian}) to achieve coherent control of spin state. To precisely determine the relevant parameters \cite{SI}, we perform Ramsey measurement and determine the frequency detuning $\Delta_z$, see Fig.\ref{fig:setup}(c). Subsequently, we measure Rabi oscillation and estimate the effective Rabi frequency $\omega_0=\bla{\Delta_x^2+\Delta_z^2}^{1/2}$, see Fig.\ref{fig:setup}(d), from which we are able to determine $\Delta_x$. We note that $\omega_0$ quantifies the energy gap between the eigenstates of $H_s=(\Delta_z/2)\sigma_z+(\Delta_x/2)\sigma_x$. We are interested in the weak driving with a large detuning so that no evident conventional Rabi oscillation is expected. We remark that the Floquet Hamiltonian Eq.(\ref{eq:Hamiltonian}) is also feasible by applying a magnetic field that has both longitudinal and transversal components (namely $\Delta_z$ and $\Delta_x$).

According to Floquet theory, the evolution dynamics of the periodically driven two-level system can be described by Floquet quasi energy states (namely Floquet modes) $\ket{\phi_{\sigma}(t)}$ with the corresponding quasienergy $\epsilon_{\sigma}$ as $\ket{\Psi(t)}=e^{-i\epsilon_\sigma t} \ket{\phi_\sigma(t)}$. The Floquet states and spectrum were observed in a strongly driven superconducting flux qubit \cite{Deng15}. The Floquet dynamics of a weakly driven two-level system can be mapped to a two-band Wannier-Stark ladder model \cite{Mendez81,Grosso2000,Russomanno17}. The upper and lower bands arising from Floquet synthetic dimensions \cite{Martin17} correspond to the spin states $\ket{\pm}$ that are the eigenstates of $H_s$ \cite{SI}, namely  $\ket{+}=\cos(\frac{\theta}{2})\ket{0}+\sin(\frac{\theta}{2})\ket{-1}$ and $\ket{-}=-\sin(\frac{\theta}{2})\ket{0}+\cos(\frac{\theta}{2})\ket{-1}$ with  $\theta=\tan^{-1}\bla{\Delta_x/\Delta_z}$. The energies of the upper and lower Floquet levels denoted as $\ket{\pm,n}$ are $E_{\pm,n}=\pm( \omega_0/2) +n \omega$, where $\pm$ represents the upper and lower band respectively, see Fig.\ref{fig:frt}(a). As $\ket{\pm}$ are not the eigenstates of $\sigma_x$, the effect of the weak driving field as written in the Hamiltonian Eq.(\ref{eq:Hamiltonian}) will couple the levels $\ket{\alpha,n}$ with $\ket{\beta,n\pm1}$, where $\alpha,\beta=\pm$, see Fig.\ref{fig:frt}(a). The strength of such off-resonant coupling is $(A/2\omega_0)\blb{\delta_{\alpha\beta}\Delta_x+ (1-\delta_{\alpha\beta})\Delta_z}$ \cite{Russomanno17,SI}. Therefore, it can be seen that the transition between the upper and lower band as meditated by Floquet levels can be exploited to manipulate spin state in a coherent manner.

In our experiment, we exploit the three-level configurations provided by Floquet synthetic levels, allowing the implementation of Raman transitions between the upper and lower levels. For example, $\ket{+,n-1}$, $\ket{-,n+1}$ and $\ket{+,n}$ ($\ket{-,n}$) forms a $\Lambda$ (V) three-level configuration, see Fig.\ref{fig:frt}(b). In order to enable efficient far-detuned Raman transition between $\ket{+,n-1}$ and $\ket{-,n+1}$ via the intermediate states $\ket{\pm,n}$, the energy resonant condition has to be satisfied, namely $E_{+,n-1}=E_{-,n+1}$ \cite{Russomanno17,SI}. This leads to the Floquet resonance condition $\omega=(\omega_0/2)$, which contrasts to the conventional Rabi resonance condition $\omega=\omega_0$ for a weakly driven two-level system \cite{Russomanno17}. The resonance conditions are slightly modified due to higher-order dynamical Stark shifts \cite{Russomanno17}. We first prepare the NV center spin in the state $\ket{+}$, namely in the upper band, by applying a microwave pulse on resonance with the two-level system with the amplitude $\Omega$ for a time duration $\tau_{\theta}=\theta/(2\Omega)$ to induce a rotation of angle $\theta$ around $\hat{y}$ axis. The system is then governed by the Floquet Hamiltonian as in Eq.(\ref{eq:Hamiltonian}) arising from the microwave driving field $V(t)$, see Fig.\ref{fig:setup}(b). We measure the state population of the spin level $\ket{0}$ as a function of the evolution time. As shown in Fig.\ref{fig:frt}(c), our experimental data clearly demonstrates Floquet Raman transition between the upper and lower band. Under the resonant condition, the state population $P_{\ket{0}}$ can be written as
$P_{\ket{0}}=\frac{1}{2}\blb{1+\cos\theta\cos\bla{\Omega_F t}-\sin\theta\sin\bla{\Omega_F t}\sin\bla{\omega_0 t}}$ \cite{SI}, where $\Omega_F$ is the Rabi frequency induced by Floquet Raman transition. We remark that the additional fast oscillation feature is mainly due to the interband energy splitting $\omega_0$, and can be eliminated by an additional appropriate microwave pulse \cite{SI}. We further characterize the dependence of the Rabi frequency of the Floquet Raman transition on the driving strength $A$, see Fig.\ref{fig:frt}(d), which agrees well with our numerical simulation and the theoretical analysis taking into account the Bloch-Siegert-like effect \cite{Saiko2016,Saiko2018,SI,Russomanno17}. In Fig.\ref{fig:frt}(e), we show the relative contrast of Floquet Raman transition which clearly demonstrates Floquet resonant feature, namely an optimal Raman transition efficiency appears when the resonant condition is satisfied.
%

%
\begin{figure}[t]
\begin{center}
\hspace{-0.5cm}
\includegraphics[width=9cm]{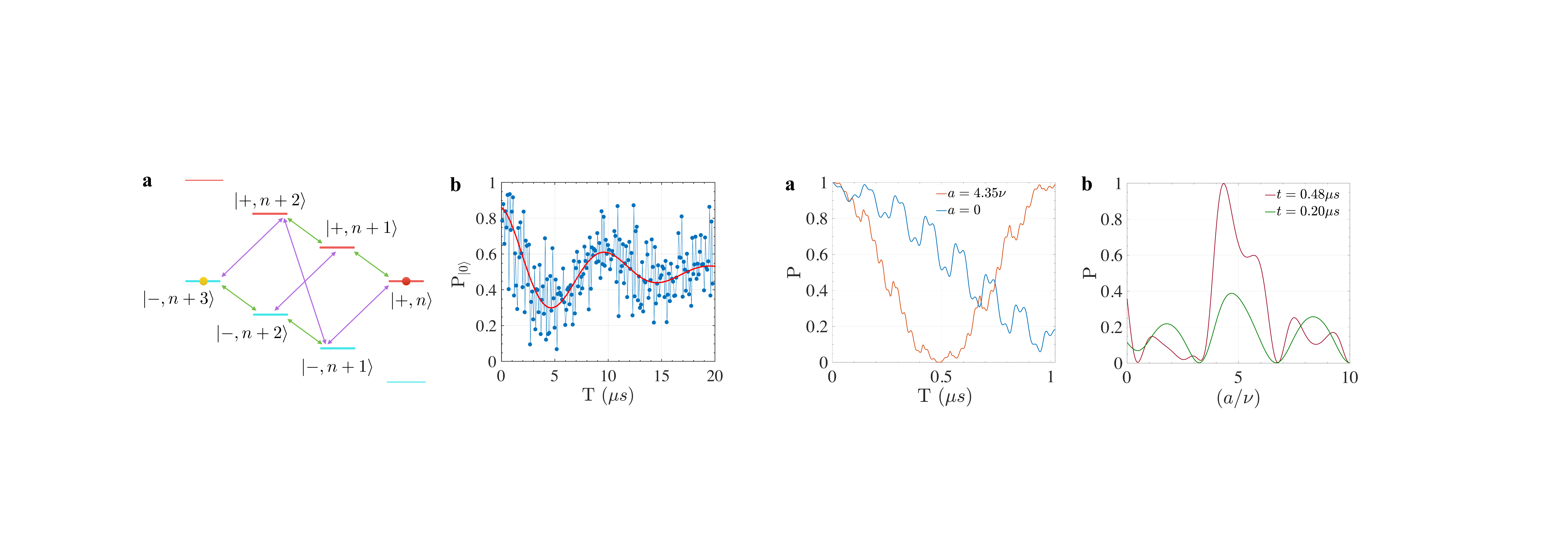}
\end{center}
\caption{(Color online) Third-order Floquet Raman transition between the lower and upper band levels $\ket{-,n+3}$ and  $\ket{+,n}$. (a) The Raman transition is mediated by multiple intermediate levels $\ket{\pm,n+1}$ and $\ket{\pm,n+2}$. (b) The state $\ket{0}$ population as a function of time demonstrates coherent Raman Floquet transition of the spin state as realized by a third-order Flquet resonance. The red curve represents inter-band transition filtering out the fast oscillating component. The frequency of the weak microwave driving field is $\omega=(2\pi)  4.657$ MHz, and the other parameters are $A=(2\pi)  2.37$ MHz, $\Delta_z=(2\pi) 9.82$ MHz, $\Delta_x=(2\pi) 9.67 $ MHz.}
\label{fig:third_order}
\end{figure}
%

Raman transition between the other Floquet synthetic levels are also feasible if the general $m$-th order Floquet resonance condition $\omega=(\omega_0/m) $ is satisfied \cite{Russomanno17}, where Floquet Raman transition is mediated by multiple levels \cite{Cohen_book,Bateman10}, see Fig.\ref{fig:third_order}(a). In our experiment, we tune the frequency of the weak microwave driving field to match the third-order Floquet resonance condition $\omega= (\omega_0/3)$. The coherent oscillation of the spin state population as shown in Fig.\ref{fig:third_order}(b) demonstrates the third-order Floquet resonance enabled Raman transition. 
The observed Floquet Raman transition allows to coherently manipulate spin state using a driving field with limited frequency that is a fraction of the one in conventional Rabi resonance, and thus may facilitate coherent quantum control e.g. of spin system under a high magnetic field. 

We remark that the experiment parameters satisfy the traditional adiabatic condition \cite{SI}, which implies that the system is expected to remain in the initial spin eigenstate \cite{Marzlin04,Tong10,Tong05}. Instead, our experiment observation demonstrates evident spin state transition (namely deviating from the adiabatic behavior). The phenomenon  represents more general anomalous non-adiabaticities in weakly driven systems  \cite{Du2008PRL}.

\begin{figure}[t]
\begin{center}
\hspace{-0.2cm}
\includegraphics[width=9cm]{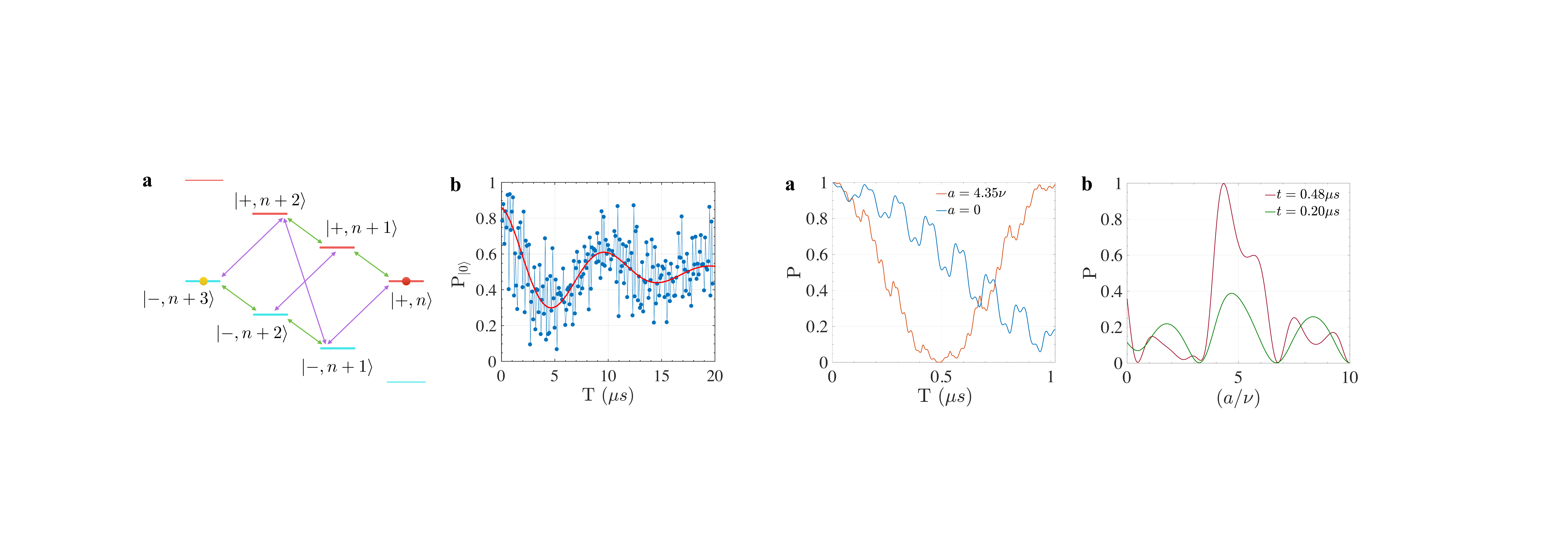}
\end{center}
\caption{(Color online) Theoretical prediction of analog photon-assisted Floquet Raman transition. ({\bf a}) The state probability $P$ in the upper band as a function of time with ($a=4.35\nu$) and without ($a=0$) frequency modulation. The system is initialized in the upper band. ({\bf b}) The state probability $P$ in the lower band at time $T=0.20$ $\mu s$ and $0.48$ $\mu s$ as a function of the phase modulation parameter $(a/\nu)$. The other parameters are  $\omega=(2\pi)  7.271$ MHz, $A=(2\pi)  2.404 $ MHz, $\nu=(2\pi)  7.343$ MHz, and  $\Delta_z=\Delta_x=(2\pi) 10$ MHz.}
\label{fig:photon-assisted-FRT}
\end{figure}

The observed Floquet-Raman transition allows effective tunnelling between the upper and lower band. As inspired by photon-assisted tunnelling \cite{Eckardt2005,Sias2008} and resonance enhanced quantum transport \cite{Vaziri2010}, one can show that a phase-modulated driving field $V_d(t)=A\cos\blb{\omega t + (a/\nu)\sin(\nu t)}$ leads to delocalisation in a tilted Wannier-Stark ladder \cite{SI}. Such a phenomenon together with Floquet-Raman transition makes it feasible to observe non-trivial effects in a two-band Wannier-Stark ladder as simulated by a weakly driven two-level system. As a consequence, a weak phase-modulated driving field may enhance the speed of Floquet-Raman transition, see Fig.\ref {fig:photon-assisted-FRT}(a). The system demonstrates both enhanced tunnelling and dynamical localisation \cite{Eckardt2005,Sias2008,Dunlap,Vaziri2010}, see Fig.\ref {fig:photon-assisted-FRT}(b), which has connections with the properties of Bessel function. The result makes a step towards the exploration of interesting phenomena in a two-band Wannier-Stark ladder model as simulated by a weakly driven two-level system \cite{SI}.
%

%
%

To summary, we simulate a two-band Wannier-Stark ladder model using a solid state spin system under weak periodic driving. We exploit the synthetic dimensions in the Floquet framework and experimentally observe second- and third-order Raman transition meditated by Floquet synthetic levels.  We show that analog photon-assisted Floquet Raman transition may exist in such a driven two-level quantum system. The present result enriches Floquet dynamics for Floquet Hamiltonian engineering aiming for the development of new techniques for coherent quantum control and the investigation of novel Floquet quantum phases. In addition, the observed Floquet Raman transition provides additional and more general scenarios which violate the widely adopted traditional adiabatic condition, and may provide insights into adiabatic dynamical evolution of quantum systems.

{\em Acknowledgements --} We thank Dr. A. Russomanno, Dr. H.-B. Liu and Dr. J.-F. Wang for helpful discussions. The work is supported by National Natural Science Foundation of China (11690030, 11690032, 11574103), National 1000 Youth Talent Program, Huazhong University of Science and Technology. M. B. P. is supported by the DFG (FOR1493), the EU STREPs DIADEMS and HYPERDIAMOND, and the ERC Synergy grant BioQ. F. J. acknowledge support from DFG (FOR 1493, SPP 1923), VW Stiftung, BMBF, ERC, EU (DIADEMS), BW Stiftung, Ministry of Science and Arts, Center for Integrated quantum science and technology (IQST).



\section*{Supplementary material}

# Supplementary Information

## 1. Derivation of the effective Hamiltonian

By applying a microwave driving field as follows

$$V(t) = \Delta_x \cos(\omega_d t) + 2A \cos(\omega_d t) \sin(\omega t), \tag{S.1}$$

the system Hamiltonian in the the lab frame can be written as follows

$$H_o = \frac{\Delta E}{2}\sigma_z + \Delta_x \cos(\omega_d t)\sigma_x + 2A\cos(\omega_d t)\sin(\omega t)\sigma_x, \tag{S.2}$$

where $\Delta E$ is the energy of the two-level system. Here, the microwave field is designed to drive the transition between the spin sublevels $|0\rangle$ and $|-1\rangle$, while the other level $|+1\rangle$ is far detuned. Therefore, we can concentrate on the Hilbert subspace as spanned by $\{|0\rangle, |-1\rangle\}$, and $\sigma_x$ and $\sigma_z$ are Pauli operators of such an effective two-level system. In the interaction picture with respect to $H_0 = (\omega_d/2)\,\sigma_z$, we get the following Hamiltonian as

$$H_I = \frac{\Delta_z}{2}\sigma_z + \frac{\Delta_x}{2}\sigma_x + A\sin(\omega t)\,\sigma_x \tag{S.3}$$

$$+ \frac{\Delta_x}{2}\left[\cos(2\omega_d t)\,\sigma_x - \sin(2\omega_d t)\,\sigma_y\right] + A\left[\cos(2\omega_d t)\,\sigma_x - \sin(2\omega_d t)\,\sigma_y\right]\sin(\omega t), \tag{S.4}$$

where $\Delta_z = \Delta E - \omega_d$. In our experiment, the condition $\omega, A, \Delta_x \ll \omega_d$ is satisfied. Therefore, under rotating wave approximation, the last two fast oscillating terms in Eq.(S.4) can be neglected, and we get the following effective Hamiltonian as implemented in the experiment

$$H = \frac{\Delta_z}{2}\sigma_z + \frac{\Delta_x}{2}\sigma_x + A\sin(\omega t)\sigma_x, \tag{S.5}$$

that allows us to observe Floquet Raman transition in the synthetic dimensions.

## 2. Determination of experiment parameters

In our experiments, we determine the NV center spin state from fluorescence data. The data shown in Figure 1 (c-d), Figure 2(c) and Figure 3(b) of the main text represents the normalized florescence data in units of $m_s = 0$ state probability $P_{|0\rangle}$ , where $P_{|0\rangle} = 1$ and $P_{|0\rangle} = 0$ correspond to the maximum and the minimum fluorescence respectively

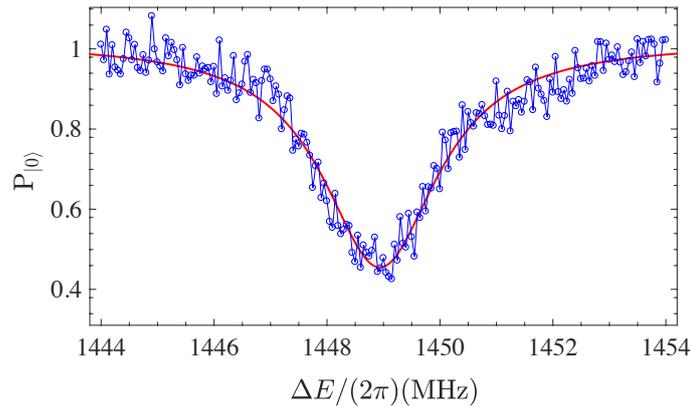

Figure S1: (Color online) The measurement of pulsed ODMR experiment gives an estimation of the resonance frequency 1448.9MHz, which corresponds a magnetic field around 508Gauss in parallel with the NV axis. The microwave pulse length is 500 ns with a Rabi frequency $\Omega \simeq (2\pi)1$ MHz.

[1, 2]. In order to determine the resonance frequency accurately, we first perform pulsed optically detected magnetic resonance (pulsed ODMR) measurement. The pulsed ODMR experiment, as shown in Fig.S1, exhibits one single resonant frequency with a finite linewidth, which confirms the polarization of the nitrogen nuclear spin associated with the NV center. To determine the frequency detuning $\Delta_z$ of the microwave driving field, we perform Ramsey experiment and determine the value of $\Delta_z$ from the oscillation frequency of Ramsey fringe, as shown in Fig.1(c) of the main text. The value of $\Delta_x$ is determined from the measurement of Rabi oscillation. Given a frequency detuning of $\Delta_z$, the effective Rabi frequency is

$$\omega_0 = (\Delta_x^2 + \Delta_z^2)^{1/2}, \tag{S.6}$$

from which we can infer the value of the parameter $\Delta_x$.

### 3. Mapping between a periodically driven system to a Wannier-Stark ladder model

The presence of Floquet resonance in the present model was proposed in Ref.[3]. In this section, we provide theoretical analysis to elaborate our experiment observation of Floquet Raman transition. The evolution of a periodically driven system can be described in the framework of Floquet theory as

$$|\Psi_\alpha(t)\rangle = e^{-i\epsilon_\alpha t}|\phi_\alpha(t)\rangle, \tag{S.7}$$

where the Floquet state $|\phi_\alpha(t)\rangle = \phi_\alpha(t+T)$ is periodic in time with quasienergy $\epsilon_\alpha$. Considering a two-level system as governed by the Hamiltonian in Eq.(S.5), the Floquet state can be expanded as

$$|\phi_\alpha(t)\rangle = \sum_{n=-\infty}^{\infty} c_{\alpha,n,0}|0\rangle e^{in\omega t} + \sum_{n=-\infty}^{\infty} c_{\alpha,n,-1}|-1\rangle e^{in\omega t} \tag{S.8}$$

its dynamical evolution is governed by equation

$$\left\{ H(t) - i\frac{\partial}{\partial t} \right\} |\phi_\alpha(t)\rangle = \epsilon_\alpha |\phi_\alpha(t)\rangle. \tag{S.9}$$

Define $|n, 0\rangle = |0\rangle e^{in\omega t}$ and $|n, -1\rangle = |-1\rangle e^{in\omega t}$, the matrix form of the Floquet Hamiltonian $\mathcal{H}(t) = H(t) - i\frac{\partial}{\partial t}$ can be written as

$$\mathcal{H} = \sum_{n,\sigma=0,-1} \left( n\omega + \frac{\Delta_z}{2}\sigma_z \right)|n,\sigma\rangle\langle n,\sigma| + \frac{\Delta_x}{2}\sum_n \left( |n,0\rangle\langle n,-1| + h.c. \right) + \frac{A}{2}\sum_n \left( |n,0\rangle\langle n+1,-1| + |n,0\rangle\langle n-1,-1| + h.c. \right), \tag{S.10}$$

here we use a translation $t \to t + \pi/(2\omega)$ in Eq.(S.5) so that $\sin(\omega t) \to \cos(\omega t)$. We remark that such a translation is only for the simplicity of derivation and would not affect the physics. We denote $H_s = (\Delta_z/2)\sigma_z + (\Delta_x/2)\sigma_x$, which has the following eigenstates

$$|+\rangle = \cos\left(\frac{\theta}{2}\right)|0\rangle + \sin\left(\frac{\theta}{2}\right)|-1\rangle, \tag{S.11}$$

$$|-\rangle = -\sin\left(\frac{\theta}{2}\right)|0\rangle + \cos\left(\frac{\theta}{2}\right)|-1\rangle. \tag{S.12}$$

The Floquet state can be rewritten in this new basis and the corresponding matrix form of Floquet Hamiltonian will be rewritten as

$$\begin{aligned}
\mathcal{H} = &\sum_{n,\sigma=\pm} \left( n\omega + \frac{\omega_0}{2}\sigma_z \right)|n,\sigma\rangle\langle\sigma,n| + \frac{A\Delta_x}{2\omega_0}\sum_n \left( |n,+\rangle\langle n+1,+| - |n,-\rangle\langle n+1,-| + h.c. \right) \\
&+ \frac{A\Delta_z}{2\omega_0}\sum_n \left( |n,+\rangle\langle n+1,-| + |n,-\rangle\langle n+1,+| + h.c. \right).
\end{aligned} \tag{S.13}$$

where $\omega_0 = \sqrt{\Delta_x^2 + \Delta_z^2}$ and the basis $|n,\pm\rangle = |\pm\rangle e^{in\omega t}$. It is a two-leg Wannier-Stark ladder model with a tilted field $\omega$ [4].

## 4. Theoretical analysis of Floquet-Raman transition

The energies of the Floquet synthetic levels $|n, \pm\rangle$ (as shown in Fig.2a of the main text) are $E_{\pm,n} = \pm(\omega_0/2) + n\omega$. The second and third terms in Eq.(S.13) shows the coupling between the Floquet synthetic levels. If $\Delta_z = 0$, the interleg tunneling can be ignored, the system can be considered as two independent one-leg Wannier-Stark ladders. These two ladders can be diagonalized as follows

$$\mathcal{H}_{\text{intra}} = \sum_{m,\sigma=\pm} \left(\frac{\omega_0}{2}\sigma_z + m\omega\right)|\Psi_{m,\sigma}\rangle\langle\Psi_{m,\sigma}|, \tag{S.14}$$

where

$$|\Psi_{m,+}\rangle = \sum_n \mathcal{J}_{m-n}\left(\frac{A\Delta_x}{\omega_0\omega}\right)|n,+\rangle \qquad |n,+\rangle = \sum_m \mathcal{J}_{m-n}\left(\frac{A\Delta_x}{\omega_0\omega}\right)|\Psi_{m,+}\rangle$$

$$|\Psi_{m,-}\rangle = \sum_n \mathcal{J}_{n-m}\left(\frac{A\Delta_x}{\omega_0\omega}\right)|n,-\rangle \qquad |n,-\rangle = \sum_m \mathcal{J}_{n-m}\left(\frac{A\Delta_x}{\omega_0\omega}\right)|\Psi_{m,-}\rangle. \tag{S.15}$$

With the new basis, the interleg Hamiltonian can be expanded as

$$\begin{aligned}\mathcal{H}_{\text{inter}} &= \frac{A\Delta_z}{2\omega_0}\sum_n \left(|n,+\rangle\langle n+1,-| + |n,-\rangle\langle n+1,+| + h.c.\right)\\ &= \frac{\Delta_z\omega}{2\Delta_x}\left\{\sum_{m,m'}\left[(m-m')\mathcal{J}_{m-m'}\left(\frac{2A\Delta_x}{\omega_0\omega}\right)\right]|\Psi_{m,+}\rangle\langle\Psi_{m',-}| + h.c.\right\},\end{aligned} \tag{S.16}$$

From the effective interaction Hamiltonian (S.16), if $A \ll \omega$ and any other energy scales, the amplitude of tunneling term is very small. So the tunneling is forbidden except for the case $\epsilon_{m,+} = \epsilon_{m',-}$. Consider the dynamical process, if the initial state is prepared on the state $|0,+\rangle$, it can be expanded in new basis as

$$|0,+\rangle = \sum_m \mathcal{J}_m\left(\frac{A\Delta_x}{\omega_0\omega}\right)|\Psi_{m,+}\rangle, \tag{S.17}$$

when $A \ll \omega$, one can use the approximation $|0,+\rangle \approx |\Psi_{0,+}\rangle$.

In the following, we consider two cases of Floquet resonance.

(1). $2^{nd}$-order Floquet resonance: $\omega = \omega_0/2$. In this case, the resonance condition is $\epsilon_{m,+} = \epsilon_{m+2,-}$. The effective Hamiltonian is

$$\mathcal{H}_{\text{eff}}^{[2]} = -\frac{\Delta_z\omega}{\Delta_x}\mathcal{J}_2\left(\frac{2A\Delta_x}{\omega_0\omega}\right)\left(|\Psi_{0,+}\rangle\langle\Psi_{2,-}| + h.c.\right) \approx -\frac{A^2\Delta_z\Delta_x}{2\omega_0^2\omega}\left(|\Psi_{0,+}\rangle\langle\Psi_{2,-}| + h.c.\right) \tag{S.18}$$

In our experiment, we choose $\Delta_x \simeq \Delta_z$ to optimise the transfer efficiency, as seen from the above equation. There are other off-resonant transitions, e.g. one can extend the Hilbert space including three basis $|\Psi_{0,+}\rangle$, $|\Psi_{1,-}\rangle$ and $|\Psi_{2,-}\rangle$ as

$$\begin{aligned}\mathcal{H}_{\text{eff}}^{[2]} &= -\frac{\Delta_z\omega}{\Delta_x}\mathcal{J}_2\left(\frac{2A\Delta_x}{\omega_0\omega}\right)\left(|\Psi_{0,+}\rangle\langle\Psi_{2,-}| + h.c.\right) + \frac{\Delta_z\omega}{2\Delta_x}\mathcal{J}_1\left(\frac{2A\Delta_x}{\omega_0\omega}\right)\left(|\Psi_{0,+}\rangle\langle\Psi_{1,-}| + h.c.\right) - \omega|\Psi_{1,-}\rangle\langle\Psi_{1,-}|\\ &\approx -\frac{A^2\Delta_z\Delta_x}{2\omega_0^2\omega}\left(|\Psi_{0,+}\rangle\langle\Psi_{2,-}| + h.c.\right) + \frac{A\Delta_z}{2\omega_0}\left(|\Psi_{0,+}\rangle\langle\Psi_{1,-}| + h.c.\right) - \omega|\Psi_{1,-}\rangle\langle\Psi_{1,-}|.\end{aligned} \tag{S.19}$$

The second term of this effective Hamiltonian describes off-resonant transition which may lead to fast oscillation. This effect will be suppressed when the energy detuning between two states $|\Psi_{0,+}\rangle$, $|\Psi_{1,-}\rangle$ satisfies the condition $\omega \gg \left(A\Delta_z/2\omega_0\right)$.

(2). $3^{rd}$-order Floquet resonance: $\omega = \omega_0/3$. In this case, the resonance condition is $\epsilon_{m,+} = \epsilon_{m+3,-}$. The effective Hamiltonian is

$$\mathcal{H}_{\text{eff}}^{[3]} = \frac{3\Delta_z\omega}{2\Delta_x}\mathcal{J}_3\left(\frac{2A\Delta_x}{\omega_0\omega}\right)\left(|\Psi_{0,+}\rangle\langle\Psi_{3,-}| + h.c.\right) \approx \frac{A^3\Delta_z\Delta_x^2}{4\omega_0^3\omega^2}\left(|\Psi_{0,+}\rangle\langle\Psi_{3,-}| + h.c.\right) \tag{S.20}$$

It can be seen that the transfer efficiency is optimized when $\Delta_z \simeq \Delta_x/\sqrt{2}$. In our experiment, the system is initialized into the $|+\rangle$ state, and the state evolution is described as follows

$$|\psi(t)\rangle = \exp\left(-i\sigma_x \Omega_F t/2\right)|+\rangle = \cos\left(\frac{\Omega_F}{2}t\right)|+\rangle - i\sin\left(\frac{\Omega_F}{2}t\right)|-\rangle, \tag{S.21}$$

where

$$2^{nd}\text{-order Floquet resonance:} \quad \Omega_F = \left(\frac{A^2 \Delta_z \Delta_x}{\omega_0^2 \omega}\right) = \left(\frac{2A^2 \Delta_z \Delta_x}{\omega_0^3}\right) \tag{S.22}$$

$$3^{rd}\text{-order Floquet resonance:} \quad \Omega_F = \left(\frac{A^3 \Delta_z \Delta_x^2}{2\omega_0^3 \omega^2}\right) = \left(\frac{9A^3 \Delta_z \Delta_x^2}{2\omega_0^5}\right) \tag{S.23}$$

represents the Rabi frequency of Floquet Raman transitions as see from Eq.(S.18-S.20), and $|+\rangle = \cos(\frac{\theta}{2})|0\rangle + \sin(\frac{\theta}{2})|-1\rangle$, $|-\rangle = -\sin(\frac{\theta}{2})|0\rangle + \cos(\frac{\theta}{2})|-1\rangle$, where $\theta = \tan^{-1}(\Delta_x/\Delta_z)$. Therefore, it can be seen that a complete state transfer between the spin states $|+\rangle$ and $|-\rangle$ can be achieved by applying an additional rotation around $\hat{y}$ axis, one can implement coherent transition between the states $|m_s = 0\rangle$ and $|m_s = -1\rangle$. In the lab frame, the state evolution is

$$|\psi_s(t)\rangle = \cos\left(\frac{\Omega_F}{2}t\right)\exp\left(-i\frac{\omega_0}{2}t\right)|+\rangle - i\sin\left(\frac{\Omega_F}{2}t\right)\exp\left(i\frac{\omega_0}{2}t\right)|-\rangle. \tag{S.24}$$

After the evolution for time $T$, we directly measure the population of the state $|m_s = 0\rangle$ via spin-dependent fluorescence, which can be written as follows

$$P_{|0\rangle} = |\langle 0|\psi_s(T)\rangle|^2 \tag{S.25}$$

$$= \frac{1}{2}\left[1 + \cos(\theta)\cos(\Omega_F T) - \sin(\theta)\sin(\Omega_F T)\sin(\omega_0 T)\right]. \tag{S.26}$$

The second term $\sim \sin(\omega_0 T)$ in Eq.(S.26) represent a fast oscillating feature, which agrees well with our experiment observation.

We remark that the non-RWA (rotating wave approximation) effect may lead to certain deviation between the observed and the calculated RWA Floquet Raman transition frequency. When taking into account the corrections from the Bloch-Siegert-like effect [5], the observed oscillation frequencies of the Floquet Raman transition agree well with the theoretical predictions, which is also confirmed by our numerical simulation.

### 5. Floquet Raman transition with different driving amplitudes

In our experiment, we tune the amplitude $A$ of the weak driving microwave field, and investigate the dependence of Rabi frequency of Floquet Raman transition on the driving amplitude. In Fig.S2(a-b), we show the coherent spin state transfer via Floquet Raman transition that is driven by a microwave field with different amplitude $A$. We perform numerical simulation for the evolution governed by the system Hamiltonian $H$ as in Eq.(S.5). The initial state is $|\psi(0)\rangle = |0\rangle$, and thus at time $t$ the system evolves to the following state as $|\psi(t)\rangle = \exp(-itH)|\psi(0)\rangle$. The population of the state $|0\rangle$ is obtained as $P(t)_{|0\rangle} = |\langle 0|\psi(t)\rangle|^2$. Using the parameters that we determine in the experiments, the results of our numerical simulation, as shown in Fig.S2(c-d), are in good agreement with the experimental observation.

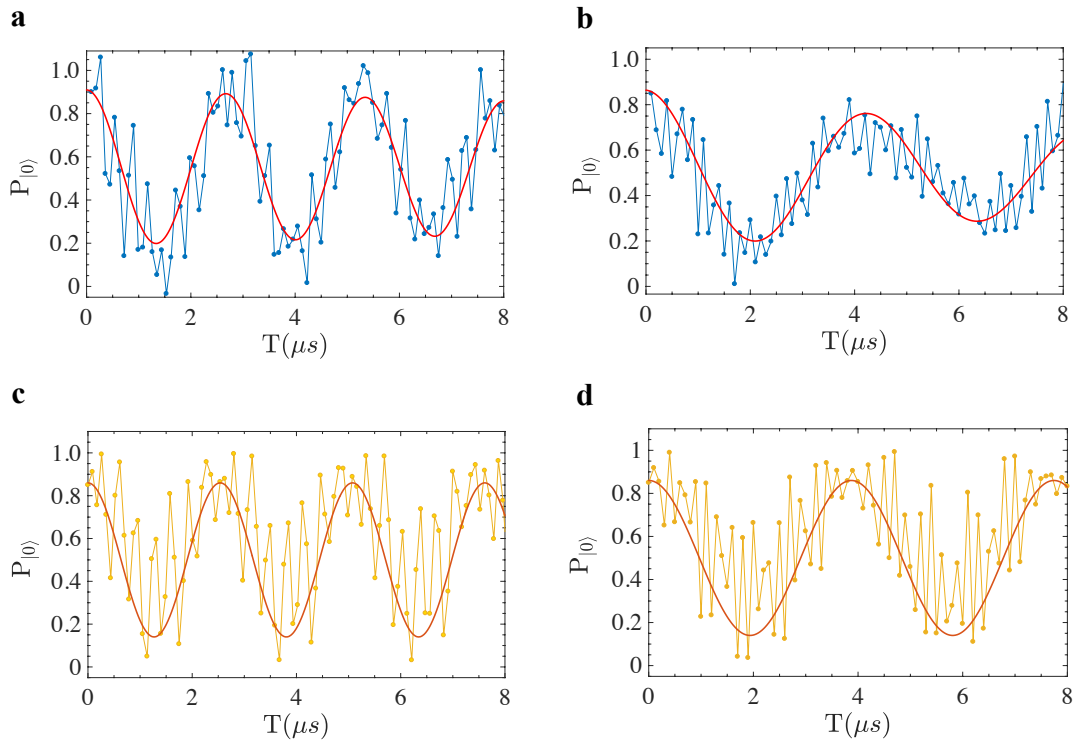

Figure S2: (Color online) Coherent spin state transfer via Floquet Raman transition that is driven by a microwave field with a frequency of $\omega$, and an amplitude of $A = (2\pi)2.22$ MHz (**a**), $(2\pi)1.79$ MHz (**b**), as compared with the numerical simulation (**a**) *vs.* (**c**) and (**b**) *vs.* (**d**). The effective Rabi frequency of Floquet Raman transition is estimated to be 0.377 MHz (**a**) *vs.* 0.386 MHz (**c**) and 0.231 MHz (**b**) *vs.* 0.240 MHz (**d**). The frequency of the driving microwave field is $\omega = (2\pi)7.09$ MHz, the other parameters are $\Delta_z = (2\pi)9.92$ MHz, and $\Delta_x = (2\pi)10.12$ MHz.

## 6. Photon-assisted Raman transition in a simple $\Lambda$ system

Periodic modulation of the trapping potential can lead to an analog of photon-assisted tunneling [6, 7], which e.g. can be implemented in sinusoidally shaken optical lattice or by applying AC electric field in solid-state systems. Here, we provide analysis and show that similar phenomena can be observed in a two-leg Wannier-Stark ladder model as simulated by a weakly driven qubit system.

To illustrate the essential idea, we first analyse the similar analog of photon-assisted phenomena in far-detuned Raman transition in a simple $\Lambda$ system with the following Hamiltonian as

$$H = \omega \left|e\right\rangle \left\langle e\right| + \mu \left(\left|e\right\rangle \left\langle g_1\right| + \left|e\right\rangle \left\langle g_2\right| + h.c.\right). \tag{S.27}$$

The transition from the state $\left|g_1\right\rangle$ to $\left|g_2\right\rangle$ represents a far-detuned Raman transition with an effective transition rate $\Omega_e = \mu^2/2\omega$, which is small as $\omega \gg \mu$. The periodic modulation of the energy levels as follows

$$V(t) = a\cos(\nu t)\left(\left|g_1\right\rangle \left\langle g_1\right| - \left|g_2\right\rangle \left\langle g_2\right|\right), \tag{S.28}$$

where $a$ and $\nu$ represents the amplitude and frequency of the modulation respectively. Define the time-dependent operator

$$K(t) = \omega t\left|e\right\rangle\left\langle e\right| + \frac{a}{\nu}\sin(\nu t)\left(\left|g_1\right\rangle\left\langle g_1\right| - \left|g_2\right\rangle\left\langle g_2\right|\right). \tag{S.29}$$

The Hamiltonian can be rotated with the unitary operator $e^{iK(t)}$ as

$$
\begin{aligned}
H_{\text{rot}}(t) =& e^{iK(t)} H(t) e^{-iK(t)} + i \frac{\partial e^{iK(t)}}{\partial t} e^{-iK(t)} \\
=& \mu \exp \left\{ i\omega t + i \left( \frac{a}{\nu} \right) \sin(\nu t) \right\} |g_1\rangle\langle e| + \mu \exp \left\{ i\omega t - i \left( \frac{a}{\nu} \right) \sin(\nu t) \right\} |g_2\rangle\langle e| + h.c. \\
=& \mu \sum_n (-1)^n \mathcal{J}_n \left( \frac{a}{\nu} \right) e^{i(\omega - n\nu)t} |g_1\rangle\langle e| + \mu \sum_n \mathcal{J}_n \left( \frac{a}{\nu} \right) e^{i(\omega - n\nu)t} |g_1\rangle\langle e| + h.c.
\end{aligned}
\tag{S.30}
$$

where $\mathcal{J}_n(x)$ is the $n$-th order Bessel function. When the resonance condition $\omega = k\nu$ is satisfied, with the rotating wave approximation, we only consider the time-independent terms in this Hamiltonian as

$$
H_{\text{eff}} = (-1)^k J |g_1\rangle\langle e| + J |g_2\rangle\langle e| + h.c.
\tag{S.31}
$$

where $J = \mu \mathcal{J}_k \left( \frac{a}{\nu} \right)$. We remark that depending on the value $\left( \frac{a}{\nu} \right)$ one obtains an enhancement in the state transfer efficiency, namely $J = \mu \mathcal{J}_k \left( \frac{a}{\nu} \right) > \Omega_e = \mu^2 / 2\omega$ as long as $\mathcal{J}_k \left( \frac{a}{\nu} \right) > \mu/\omega$, or when $\mathcal{J}_k \left( \frac{a}{\nu} \right) = 0$ one finds suppressed transport, i.e. dynamical localisation [8, 9]. If $k$ is even (odd), the amplitude of two tunneling term have the same (opposite) value. Assume the initial state is $|g_1\rangle$, the final state at time $t$ can be easily calculated as

$$
|\psi_f(t)\rangle = \frac{1 + \cos(\sqrt{2}Jt)}{2} |g_1\rangle \mp \frac{1 - \cos(\sqrt{2}Jt)}{2} |g_2\rangle - i \frac{\sin(\sqrt{2}Jt)}{\sqrt{2}} |e\rangle.
\tag{S.32}
$$

The transition probability from the state $|g_1\rangle$ to $|g_2\rangle$ is given by

$$
P(t) = \sin^4 \left( \frac{Jt}{\sqrt{2}} \right) = \sin^4 \left[ \frac{1}{\sqrt{2}} \mu \mathcal{J}_k (\frac{a}{\nu}) t \right].
\tag{S.33}
$$

At time $t = \pi/(\sqrt{2}J)$, one can get a complete state transfer from $|g_1\rangle$ to $|g_2\rangle$. The value of $k$ is even or odd does not influence the result. The above analytical results is verified by the exact numerical simulation of the dynamical evolution as governed by the system Hamiltonian in Eq.(S.27-S.28), see Fig.S3. The deviation when $a$ is small arises from the rotating wave approximation used to get the effective Hamiltonian in Eq.(S.31). The result clearly demonstrates that the photon-assisted Raman transition exhibits significantly enhanced state transfer efficiency.

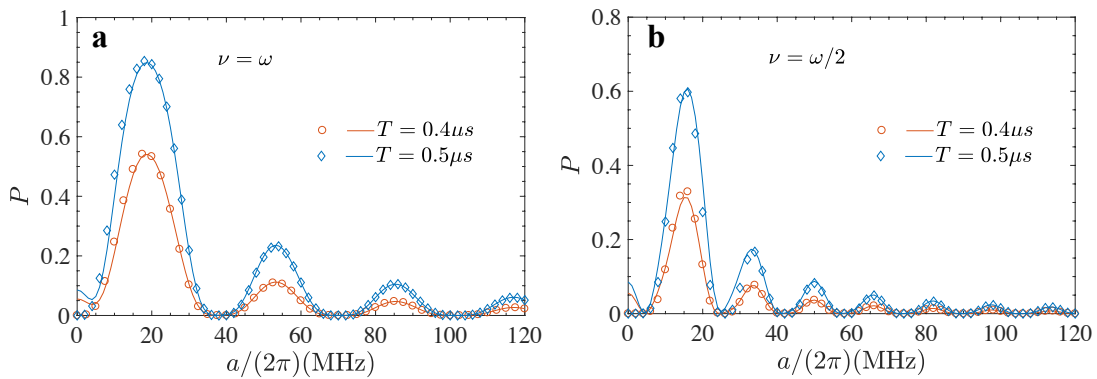

Figure S3: (Color online) The transition probability from the state $|g_1\rangle$ to $|g_2\rangle$ as a function of the amplitude of frequency modulation in a single- (a) and two- photon (b) assisted Raman processes after an evolution time of $T$. The analytical prediction ($\circ$ and $\diamond$) in Eq.(S.33) shows good agreement with the numeric simulation. The zeros of the transition probability in both plots correspond to dynamical localisation. The parameters are $\omega = (2\pi)10\text{MHz}$, $\mu = (2\pi)1\text{MHz}$.

## 7. Delocalisation in Floquet synthetic levels of a weakly driven qubit system

As inspired by the phenomenon of photon-assisted Raman transition in a simple $\Lambda$ system (see the above section), we consider the possible similar effect in a two-leg Wannier-Stark ladder model as simulated by a weakly driven qubit

system, where the driving field is periodically phase modulated as described by the following Hamiltonian

$$H(t) = \frac{\Delta_z}{2}\sigma_z + \frac{\Delta_x}{2}\sigma_x + A\cos\left\{\omega t + \frac{a}{\nu}\sin(\nu t)\right\}\sigma_x \tag{S.34}$$

If $a = 0$, the system is same with the model we discussed in main text and it can be considered as a two-leg Wannier Stark ladder. In the main text, we demonstrate that a weak driving can induce effective transitions between the upper and lower ladders. In this section, we show that the analog photon-assisted Raman transition can also appear in Floquet synthetic dimensions when $a \neq 0$, which (together the observed Floquet-Ramman transition) fascinates non-trivial effects of the synthetic two-band Wannier stark ladder model.

When $a \neq 0$, in order to better illustrate our idea, for simplicity we first consider the case $\Delta_z = 0$ with the following Hamiltonian as

$$H'(t) = \frac{\Delta_x}{2}\sigma_x + A\cos\left\{\omega t + \frac{a}{\nu}\sin(\nu t)\right\}\sigma_x = \frac{\Delta_x}{2}\sigma_x + A\sum_\ell \mathcal{J}_\ell\left(\frac{a}{\nu}\right)\cos\left\{(\omega + \ell\nu)t\right\}\sigma_x \tag{S.35}$$

where the eigenbasis of $\sigma_x$ is $|+\rangle = (|0\rangle + |-1\rangle)/\sqrt{2}$ and $|-\rangle = (|0\rangle - |-1\rangle)/\sqrt{2}$. One can find that the last time-dependent term only including the plane-waves with the frequency of sum of integer multiple of $\nu$ and $\omega$. If $\nu/\omega$ is an irrational number, the above Hamiltonian $\mathcal{H}'(t)$ can be expanded as a two-dimensional tilde lattice with the basis $|m, n, +\rangle = |+\rangle e^{i(m\omega + n\nu)t}$, the part related to $|+\rangle$ is

$$\mathcal{H}'_+ = \sum_{m,n}\left(\frac{\Delta_x}{2} + m\omega + n\nu\right)|m, n, +\rangle\langle m, n, +| + \sum_{m,n}\sum_\ell \frac{A}{2}\mathcal{J}_\ell\left(\frac{a}{\nu}\right)\left(|m, n, +\rangle\langle m + 1, n + \ell, +| + h.c.\right). \tag{S.36}$$

When two frequencies satisfy the condition $k\nu \approx \omega$, the level $|m, n, +\rangle$ almost has the same energy with the level $|m + 1, n - k, +\rangle$. As an example, we consider the simplest case of $\nu \approx \omega$. The levels $|m, n, +\rangle$ with $m + n = C$ almost has the same energy if $C$ is a constant. The levels with different $C$ have energy detuning which is much larger than their coupling, then the above two-dimensional tilted lattice can be simplified to a series of one-dimensional lattice as

$$\mathcal{H}'_{1D,+} = \sum_n\left(\frac{\Delta_x}{2} + n\delta\right)|\tilde{n}, +\rangle\langle\tilde{n}, +| + \sum_n \frac{A}{2}\mathcal{J}_\ell\left(\frac{a}{\nu}\right)\left(|\tilde{n}, +\rangle\langle\tilde{n} + 1, +| + h.c.\right) \tag{S.37}$$

where $|\tilde{n}, +\rangle = |n, -n, +\rangle$ and $\delta = \omega - \nu$ which is very small, here we only write the lattice with $C = 0$. One can choose the value of $a$ to make sure the Bessel function $\mathcal{J}_1\left(\frac{a}{\nu}\right)$ has a maximum value. It can be seen that the eigenstates of

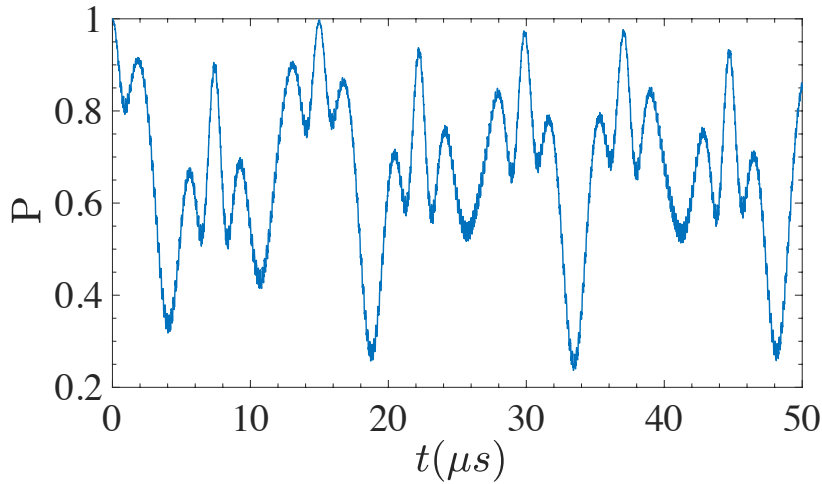

Figure S4: (Color online) The state probability in the upper band (corresponding to $|+\rangle$) as a function of time with a phase modulated driving field (Eq.S.34). The parameters are $\omega = (2\pi)7.076$ MHz, $A = (2\pi)0.5$ MHz, $\Delta_z = \Delta_x = (2\pi)10$ MHz, $\nu = (2\pi)6.942$ MHz, and $(a/\nu) = 4.8$.

this lattice can also be delocalized because of the relation $\delta \ll A$. The $|-\rangle$ component can also be expanded in the similar way as

$$\mathcal{H}'_{1D,-} = \sum_n \left( -\frac{\Delta_x}{2} + n\delta \right)|\tilde{n}, -\rangle\langle\tilde{n}, -| - \sum_n \frac{A}{2}\mathcal{J}_1\left(\frac{a}{\nu}\right)\left(|\tilde{n}, -\rangle\langle\tilde{n}+1, -| + h.c.\right). \tag{S.38}$$

Similar results can be obtained for the general $\Delta_z \neq 0$. As an example, we consider the system is initially in the state $|+\rangle$, and calculate the dynamical evolution as governed by the driven Hamiltonian $H_o$ in Eq.(S.34). In Fig.S4, it can be seen that the evolution of the state probability in the upper band (corresponding to $|+\rangle$) shows complex and rich behavior under a weak driving (rather than simple oscillation). The result hints possible non-trivial effects in a two-leg Wannier-Stark ladder model as simulated by a weakly driven two-level system.

As a short summary, we can conclude that the observed Floquet-Raman transition allows effective inter-band tunnelling between the upper and lower band, and the analog photon assisted tunnelling as discussed above allows delocalisation within individual band. Therefore, the results demonstrate that it is feasible to observe non-trivial effects in a two-leg Wannier-Stark ladder model as simulated by a weakly driven two-level system.

## 8. Violation of the traditional adiabatic condition

The adiabatic theorem states that if the Hamiltonian of a system changes slow enough, it will remains in its instantaneous eigenstate [10]. The traditional adiabatic condition is given by [11]

$$\forall m \neq n \quad \left|\frac{\langle m(t)|\dot{n}(t)\rangle}{E_m(t) - E_n(t)}\right| = \left|\frac{\langle m(t)|\dot{H}(t)|n(t)\rangle}{(E_m(t) - E_n(t))^2}\right| \ll 1, \tag{S.39}$$

with $|m(t)\rangle$ and $|n(t)\rangle$ are the instantaneous eigenstates of the time-dependent Hamilton $H(t)$, $E_m(t)$ and $E_n(t)$ are the corresponding eigenenergies. For the present Hamiltonian as in Eq.(S.5), the instantaneous eigenstates are $|e(t)\rangle = \cos(\varphi/2)|0\rangle + \sin(\varphi/2)|-1\rangle$, and $|g(t)\rangle = -\sin(\varphi/2)|0\rangle + \cos(\varphi/2)|-1\rangle$, where $\varphi = \tan^{-1}\left[(\Delta_x + 2A\sin(\omega t))/\Delta_z\right]$. The corresponding eigenenergies are $\pm\Omega_0$ with $\Omega_0 = (1/2)\left[\Delta_z^2 + (\Delta_x + 2A\sin(\omega t))^2\right]^{1/2}$. The traditional adiabatic condition Eq.(S.39) can be written as

$$\mathrm{QAC}_1 : C_1 = \left|\frac{\langle e(t)|\dot{H}(t)|g(t)\rangle}{(E_e(t) - E_g(t))^2}\right| = \left(\frac{\omega A\Delta_z}{8\Omega_0^3}\right)|\cos(\omega t)| \ll 1, \tag{S.40}$$

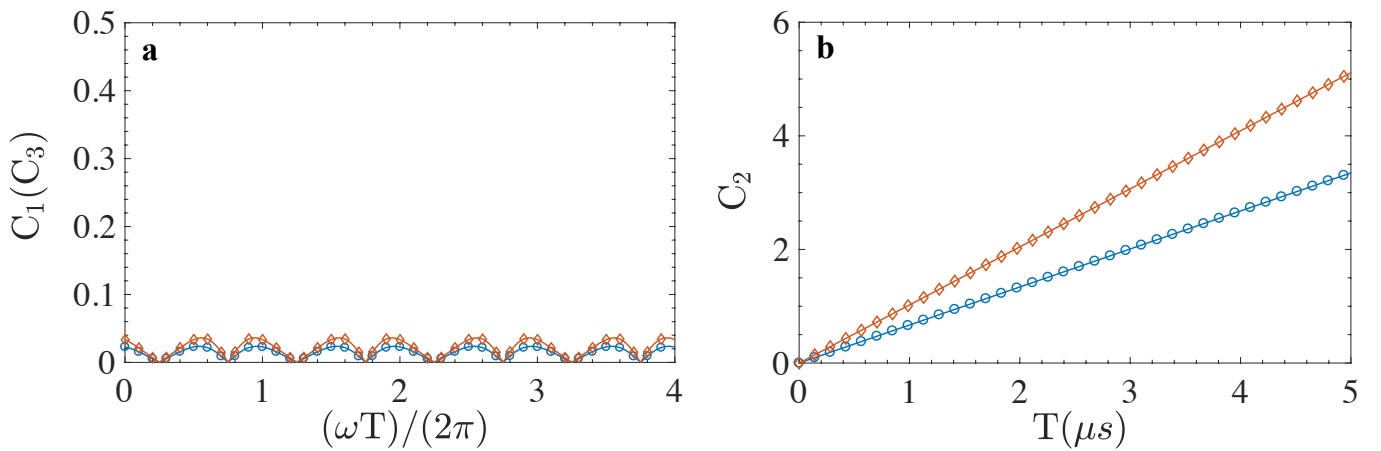

Figure S5: (Color online) The adiabatic conditions (**a**) $C_1(C_3)$ and (**b**) $C_2$ as a function of the evolution time in Floquet Raman transition. We note that C1=C3 in the present scenario.The amplitude of the microwave driving field is $A = (2\pi)0.92$ MHz (blue, ○), $(2\pi)1.35$ MHz (brown, ◇), and the frequency of the driving microwave field is $\omega = (2\pi)7.09$ MHz, the other parameters are $\Delta_z = (2\pi)9.92$ MHz, and $\Delta_x = (2\pi)10.12$ MHz.

which is neither sufficient nor necessary [12]. Ref. [13, 14] proposed a strong sufficient (but not necessary) condition to guarantee the reliability of adiabatic theorem as follows

$$\text{QAC}_2 : C_2 = \int_0^T \left| \left( \frac{\langle m(t)|\dot{n}(t)\rangle}{E_m(t) - E_n(t)} \right)' \right| dt \ll 1,\tag{S.41}$$

where $T$ is the total evolution time. Another modification of the traditional adiabatic condition is given as follows[15]

$$\text{QAC}_3 : C_3 = \left| \frac{\langle m(t)|\dot{n}(t)\rangle}{E_m(t) - E_n(t) + \Delta_{nm}(t)} \right| \ll 1,\tag{S.42}$$

and $\Delta_{nm}$ is defined as

$$\Delta_{nm}(t) = i\langle n(t)|\dot{n}(t)\rangle - i\langle m(t)|\dot{m}(t)\rangle + i\frac{d}{dt}\arg\langle m(t)|\dot{n}(t)\rangle.\tag{S.43}$$

For the present Hamiltonian Eq.(S.5) in our experiment, $\langle g(t)|\dot{g}(t)\rangle = \langle e(t)|\dot{e}(t)\rangle = 0$ and

$$\langle e(t)|\dot{g}(t)\rangle = -\frac{\dot{\varphi}}{2} = -\frac{A\Delta_z\omega\cos(\omega t)}{4\Omega_0^2}\tag{S.44}$$

is always a real number, thus $\Delta_{nm} = 0$, and the condition is equivalent to the traditional adiabatic condition, namely $C_1 = C_3$. It has been experimentally verified in [16] that $\text{QAC}_1$ is neither sufficient nor necessary, and $\text{QAC}_2$ is a strong sufficient condition, while $\text{QAC}_2$ is applicable to the system therein. We plot $C_1(C_3)$ and $C_2$ in Fig.S5 for the observed Floquet Raman transition processes. It can be seen that our experiment also demonstrates that $\text{QAC}_1$ is not sufficient, while $\text{QAC}_2$ is a strong sufficient condition, which agree with the observation in [16]. Moreover, the result instead shows that $\text{QAC}_3$ is not sufficient, see Fig.S5(a). The present experiment of Floquet Raman transition thus offers a new and more general scenario to gain insights into different formalisms of adiabatic condition [12–15, 17, 18].

---


 
 \end{document}